\documentclass[reprint,prl,aps,superscriptaddress]{revtex4-2}

\usepackage[utf8]{inputenc}

\usepackage[english]{babel}

\usepackage{color}

\usepackage{threeparttable}
\usepackage{amsfonts,amsmath,amsthm,amssymb}

\usepackage[normalem]{ulem}

\usepackage{graphicx}
\graphicspath{{img/}}
\DeclareGraphicsExtensions{.pdf,.png,.jpg}

\usepackage[colorlinks]{hyperref}

\begin{document}
\title{Observation of the metallic mosaic phase in 1\textit{T}-TaS$_2$ at equilibrium}

\author{B.~Salzmann}
\altaffiliation{Corresponding author.\\ bjoern.salzmann@unifr.ch}
\affiliation{D{\'e}partement de Physique and Fribourg Center for Nanomaterials, Universit{\'e} de Fribourg, CH-1700 Fribourg, Switzerland}

\author{E. Hujala}
\affiliation{D{\'e}partement de Physique and Fribourg Center for Nanomaterials, Universit{\'e} de Fribourg, CH-1700 Fribourg, Switzerland}
\affiliation{Lappeenranta-Lahti University of Technology LUT, FI-53850 Lappeenranta, Finland}

\author{C.~Witteveen}
\affiliation{Department of Chemistry, University of Zurich, CH-8057 Zürich, Switzerland}
\affiliation{Department of Quantum Matter Physics, University of Geneva, CH-1211 Geneva, Switzerland}

\author{B.~Hildebrand}
\affiliation{D{\'e}partement de Physique and Fribourg Center for Nanomaterials, Universit{\'e} de Fribourg, CH-1700 Fribourg, Switzerland}

\author{H.~Berger}
\affiliation{Institut de Physique des Nanostructures, {\'E}cole Polytechnique F{\'e}d{\'e}rale de Lausanne (EPFL), CH-1015 Lausanne, Switzerland}

\author{F.O.~von Rohr}
\affiliation{Department of Quantum Matter Physics, University of Geneva, CH-1211 Geneva, Switzerland}

\author{C.W.~Nicholson}
\affiliation{D{\'e}partement de Physique and Fribourg Center for Nanomaterials, Universit{\'e} de Fribourg, CH-1700 Fribourg, Switzerland}
\affiliation{Fritz Haber Institute of the Max Planck Society, D-14195, Berlin, Germany}

\author{C.~Monney}
\affiliation{D{\'e}partement de Physique and Fribourg Center for Nanomaterials, Universit{\'e} de Fribourg, CH-1700 Fribourg, Switzerland}

\date{\today}
\begin{abstract}
The transition-metal dichalcogenide tantalum disulphide (1\textit{T}-TaS$_2$) hosts a commensurate charge density wave (CCDW) at temperatures below 165~K where it also becomes insulating.
The low temperature CCDW phase can be driven into a metastable "mosaic" phase by means of either laser or voltage pulses which shows a large density of CDW domain walls as well as a closing of the electronic band gap.
The exact origins of this pulse-induced metallic mosaic are not yet fully understood.
Here, using scanning tunneling microscopy and spectroscopy (STM/STS), we observe the occurrence of such a metallic mosaic phase on the surface of TaS$_2$ without prior pulse excitation over continuous areas larger than $100 \times 100$~nm$^2$ and macroscopic areas on the millimetre scale.
We attribute the appearance of the mosaic phase to the presence of surface defects which cause the formation of the characteristic dense domain wall network.
Based on our STM measurements, we further argue how the appearance of the metallic behaviour in the mosaic phase could be explained by local stacking differences of the top layer.
Thus, we provide a potential avenue to explain the origin of the pulse induced mosaic phase.
\end{abstract}
\maketitle

\begin{figure}[tb]
\includegraphics[width=0.9\linewidth]{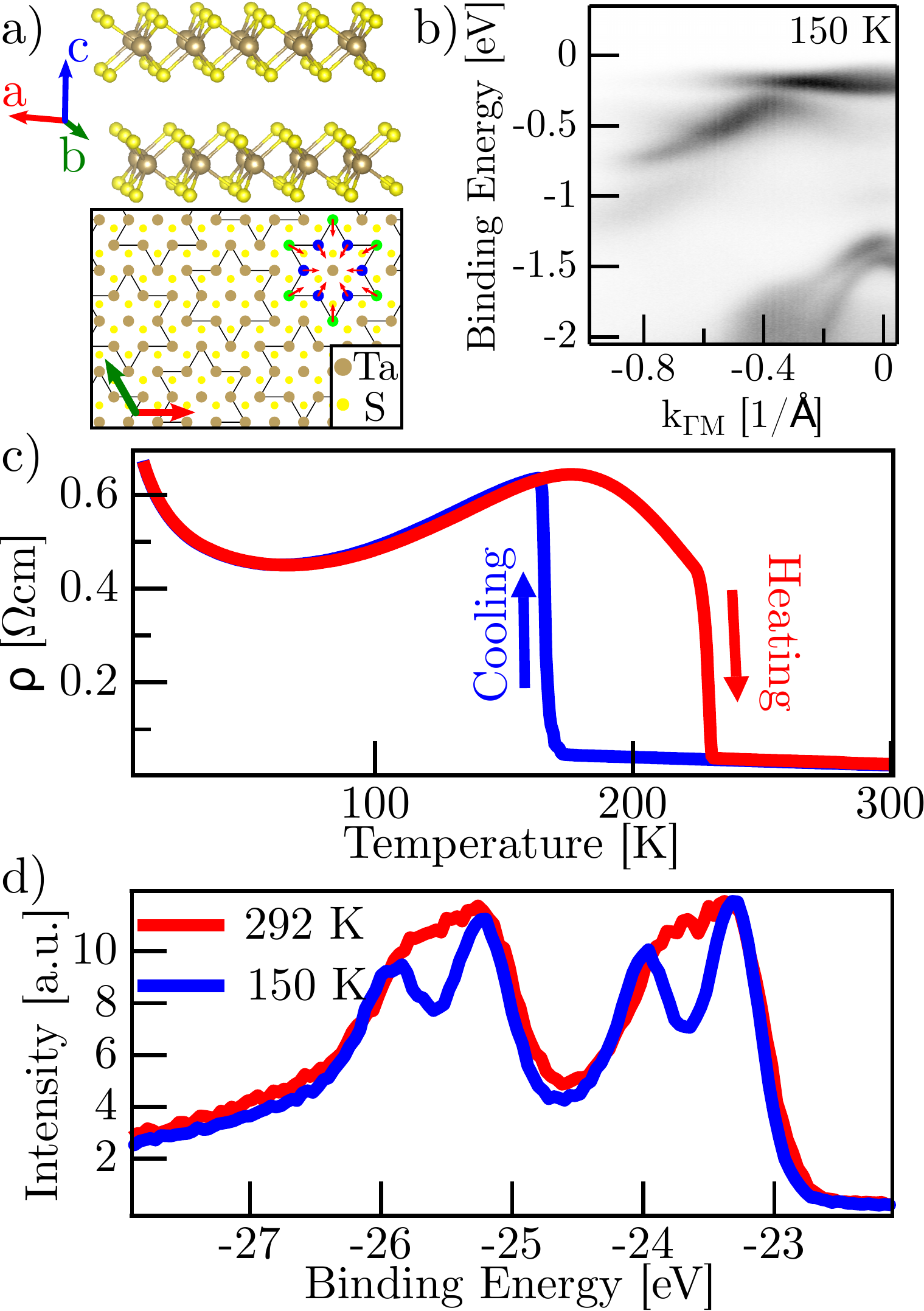} 
\caption{a) The top panel shows the atomic structure of TaS$_2$ including two layers stacked along the \textit{c} axis. The bottom panel is a view of the surface in the \textit{a}-\textit{b} plane of the material with only top layer atoms included. The SoD structure of the CDW is displayed in black across the image with the movement of the Ta atoms indicated by red arrows for one SoD. b) ARPES image taken at a photon energy of 21.2~eV at 150~K. The band structure is typical for the CCDW phase. c) Resistivity as a function of temperature measured on a TaS$_2$ sample of the same batch as the one used for other measurements in this study. The transition from the near-commensurate to the commensurate CDW phase occurs around 165~K during cooling as expected for 1\textit{T}-TaS$_2$. d) XPS measurements of a TaS$_2$ sample at different temperatures containing the Ta 4f peaks. The peak positions correspond to those expected from the 1\textit{T} polytype.}
\label{Fig:Characterisation}
\end{figure}

\textit{Introduction.} Quasi two dimensional (2D) materials present a wide range of interesting physical phenomena and promising technological applications \cite{novoselov2004,li2014,chen2018,yan2018,zhou2018}.
One class of such materials are transition metal dichalcogenides (TMDCs) which generally have a layered structure.
However, despite this structure, changes to their interlayer interactions can strongly influence their electronic properties \cite{wang2012,keum2015,wang2018,bhattacharyya2012,hong2017}.
One such material for which the influence of interlayer stacking has recently been shown to be of great relevance to the electronic properties is 1\textit{T}-TaS$_2$.
The 1\textit{T} polytype of TaS$_2$ exists in a near commensurate CDW state at room temperature with 13 Ta atoms forming a so-called star of David (SoD) pattern within the layers of the TaS$_2$ structure \cite{wu1989}.
Upon cooling, a phase transition to a fully commensurate CDW phase with a periodicity of $\sqrt{13}\times\sqrt{13}$ occurs at around 165~K, accompanied by a transition from metallic to insulating behaviour \cite{scruby1975}.
An illustration of the atomic structure of TaS$_2$ can be found in the upper panel of Figure \ref{Fig:Characterisation}a) and the in-plane CDW structure in the lower panel.
Unexpectedly, it was previously found that a so-called mosaic or hidden phase can be accessed from the CCDW phase of TaS$_2$ by applying either single laser \cite{stojchevska2014, stahl2020} or voltage pulses \cite{cho2016,hollander2015,vaskivskyi2016,ravnik2021}.
On the microscopic scale, this phase is characterised by a significantly increased density of domain walls at the site of the pulse from which its name derives.
Interestingly, within the region of these dense domain walls, the sample surface becomes metallic.
Macroscopically, a drop in resistivity can be observed accordingly.
The appearance of metallic behaviour in the normally insulating CCDW phase outside the mosaic phase has also been attributed to changes in the SoD stacking order \cite{wu2021,nicholson2022}.
SoDs can stack on top of each other in three distinct ways, as determined by the location of their center atom with respect to the other SoD: with both center atoms directly on top of each other, with the top center atom atop one of the inner six atoms of the lower SoD which are indicated in blue in the bottom panel of Figure \ref{Fig:Characterisation}a) and with the top center atom atop one of the outer six atoms, indicated in green in Figure \ref{Fig:Characterisation}a).
In the following, we will adopt the nomenclature used in ref. \cite{wu2021} and refer to these three stacking orders as  AA, AB and AC stacking respectively.
Previous investigations have determined that the bulk stacking order consists of pairs of AA stacked layers which are in turn AC stacked \cite{endo2000,hovden2016,ritschel2018,butler2020,nicholson2022,petocchi2022}.
Both AA and AC stacked layers are insulating, albeit with different origins for this insulating behaviour\cite{ma2016,butler2020,wang2020,wu2021,petocchi2022}.
On the other hand, AB stacked layers have been reported to sometimes be metallic \cite{wu2021}, presenting a potential explanation for the metallic behaviour of the mosaic phase.
However, the exact cause for the formation of the mosaic phase has so far not been established.

Here, we present STM and STS measurements of a mosaic phase occurring on an untreated surface, that is, a surface not exposed to external stimulation.
As in the pulse induced mosaic, we also observe a closing of the electronic gap within specific domains of this mosaic phase.
However, the spatial extent of the mosaic phase in our measurements is significantly larger than that of the ones induced by local stimulation and can be found across distances on the order of millimetres on the sample surface.
Based on our measurements and previous theoretical and experimental work, we suggest that the formation of the mosaic phase in our samples is caused by a high density of charged surface defects which lead to the formation of domain walls.
Through analysis of the shift between different CDW domains, we find supporting evidence that these domain walls in turn locally alter the stacking order of the top layer of the material, causing it to turn metallic.
Thus, we both explain the occurrence of a metallic mosaic in our samples and provide experimental evidence for a possible origin of the pulse induced mosaic.

\textit{Methods.}
STM measurements were taken using a commercial low temperature STM system (Scienta Omicron) at 4.5~K at a pressure of $\leq 5 \times 10^{-11}$~mbar using the constant current method with a current of 0.1~nA unless otherwise noted.
STS measurements were carried out using a lock-in technique with a frequency of 955~Hz.
Samples were cleaved using scotch tape at room temperature at a pressure of $\leq 5 \times 10^{-8}$~mbar before being transferred to the already cold STM or other measurement chambers.
Angle resolved photoemission spectroscopy (ARPES) measurements were obtained with the 21.2~eV photon energy He I spectral line produced by a commercial UV lamp (Specs GmbH) and a commercial hemispherical electron analyser (Scienta Omicron).
X-ray photoemission spectroscopy (XPS) measurements were performed using a photon energy of 1486.6~eV, produced from monochromatised Al K-$\alpha$ emission (Specs GmbH).
Samples which showed the mosaic phase in STM measurements generally did not undergo other prior measurements.

\textit{Results.}
To verify that our samples do not show any macroscopic differences to previously studied 1\textit{T}-TaS$_2$ samples and that our findings are therefore generally applicable to 1\textit{T}-TaS$_2$ samples, we have performed several characterisation measurements, the results of which are presented in figures \ref{Fig:Characterisation}b)-d).
Figure \ref{Fig:Characterisation}b) shows a representative ARPES image obtained with a photon energy of 21.2~eV at a temperature of 150~K, along the $\bar{\Gamma}\bar{\textrm{M}}$ direction of the surface Brioullin zone.
The dispersion is in good agreement with previous ARPES measurements of 1\textit{T}-TaS$_2$ in the CCDW phase \cite{perfetti2005,ngankeu2017,wang2020}.
Figure \ref{Fig:Characterisation}c) display XPS spectra of the Ta 4f core levels at room temperature and 150~K i.e. above and below the CCDW phase transition temperature.
The two peak structure is expected due to the CDW creating two inequivalent Ta sites on the outer and inner ring of the SoD structure \cite{hughes1995}.
The enhanced peak splitting observed below the CCDW transition temperature is ascribed to a stronger CDW amplitude after the transition \cite{hughes1995}.
Overall, these findings are in agreement with previous XPS measurements \cite{hughes1995,hellmann2010}.
Figure \ref{Fig:Characterisation}d) presents resistivity as a function of temperature.
The jump in resistivity occurs at the near commensurate to commensurate phase transition at around 165~K upon cooling as expected from previous studies \cite{sipos2008, wang2020} and agrees with observations from temperature dependent ARPES measurements in which we find a similar transition temperature.
Additional characterisation measurements can be found in the supplementary material and also show no unusual behaviour.
In summary, we find that our samples macroscopically behave as expected from previous literature using the above characterisation techniques.

\begin{figure}[tb]
\includegraphics[width=0.9\linewidth]{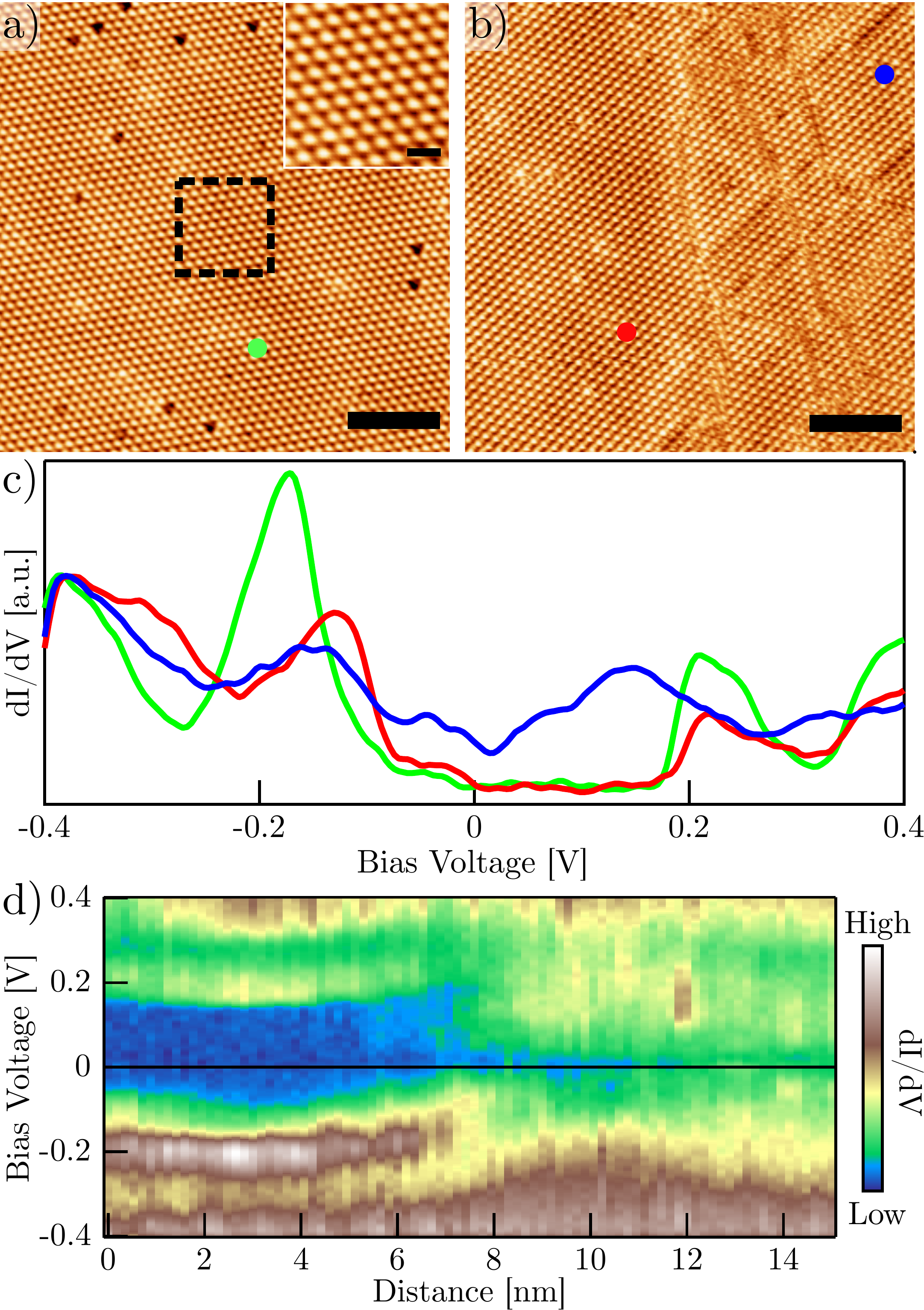} 
\caption{a) Typical STM image taken in the insulating CCDW phase of 1\textit{T}-TaS$_2$. Inset: An enlarged, defect free section of the main image marked by the dashed square therein. Note that the bright spots forming the lattice correspond to SoDs, not individual atoms. The scale bar has a length of 10~nm for the main image and 4~nm for the inset. b) STM image of a typical transition between the insulating CCDW phase on the left and the metallic mosaic phase on the right accompanied by a slight visual contrast. The image was taken at a different location on the same cleave as a). The scale bar has a size of 10~nm. c) Three STS spectra obtained at the points marked with the corresponding color in a) and b). The normal insulating behaviour occurs in the non-mosaic parts of both images while the mosaic phase is metallic. d) Cut across an STS map of a transition between insulating and metallic regions across a domain wall.}
\label{Fig:SpatialOverview}
\end{figure}

Figure \ref{Fig:SpatialOverview}a) shows an STM image of the surface of TaS$_2$ taken with a gap voltage of -0.4~V.
The inset presents a magnified section of the main image, indicated by the dashed square.
In both the main image and the inset, the bright spots forming the periodic lattice correspond to individual SoDs in the CDW phase and not to individual surface atoms.
The appearance of the surface is as expected from previous observations on 1\textit{T}-TaS$_2$ in the CCDW phase with the SoDs aligned in a hexagonal lattice.
However, our images contain an increased number of dark defect sites as compared to literature which we will discuss later.
In contrast, Figure \ref{Fig:SpatialOverview}b) is another image of the TaS$_2$ surface taken on the same cleave surface as Figure \ref{Fig:SpatialOverview}a) without any intervening treatment of the sample and measured within hours of the image in Figure \ref{Fig:SpatialOverview}a).
Notably, the surface on the right side of the image is filled with a dense network of domain walls, as seen in the pulse-induced mosaic phase previously described in TaS$_2$ \cite{stojchevska2014, ma2016, cho2016}.
Additionally, there appears to be a variation in intensity at or near the transition between the regular and mosaic regions of the image.
This contrast is again strongly reminiscent of both the pulse induced mosaic phase, see for example Figure 2 in Ref. \cite{ma2016}.

In Figure \ref{Fig:SpatialOverview}c) we compare STS spectra obtained at three different points in Figures \ref{Fig:SpatialOverview}a) and b) as marked by dots of the corresponding color.
While the spectra taken at points on the non-mosaic part of the surface are gapped as expected for the CCDW phase, the blue spectrum taken within a domain of the mosaic region is metallic, as for the pulse induced mosaic.
Further STS measurements, presented in the supplementary materials, obtained at different locations in Figure \ref{Fig:SpatialOverview}b) confirm that the visual contrast between the normal and mosaic phase consistently coincides with a change in the electronic behaviour from insulating to metallic.
To further verify whether this naturally occurring mosaic behaves the same electronically as the pulse induced mosaic, Figure \ref{Fig:SpatialOverview}d) contains a cut across an STS map taken across the boundary between the normal and metallic mosaic phases which agrees very well with data previously obtained for the pulse induced mosaic phase \cite{ma2016}.
In particular, the shifting of the peak at -0.2~V towards the Fermi level and an accompanying shift of the gap into the unoccupied states is very similar to the pulse induced mosaic phase.
 
\begin{figure*}
\includegraphics[width=0.8\linewidth]{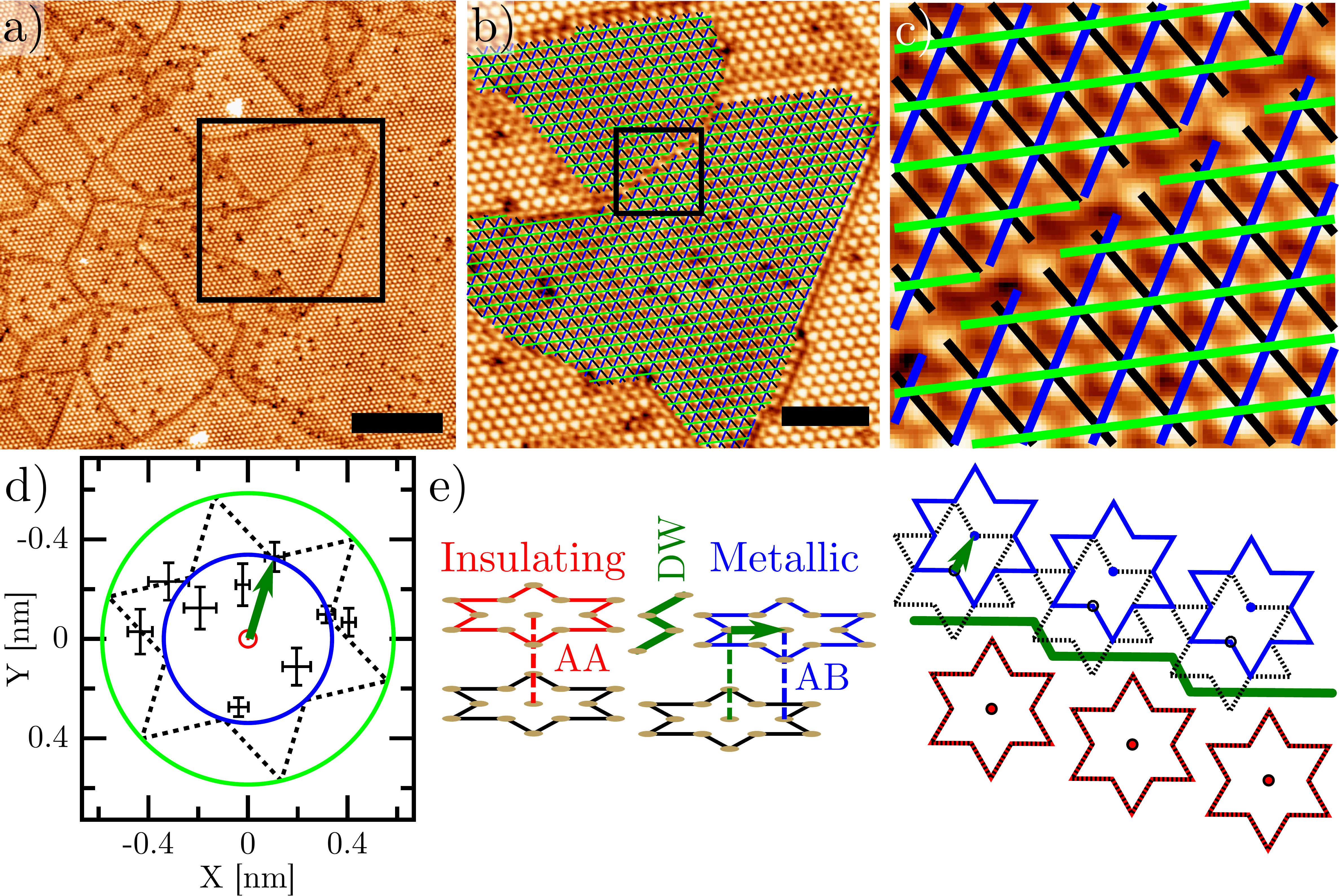} 
\caption{a) STM image of a region containing both metallic and insulating domains, as indicated by their differing intensities. The scalebar has a size of 20~nm. b) Zoom on the region marked in b) with lines fitted to SoDs in two adjacent domains of different electronic character. The resulting crossing points align well with the SoD lattice. The scalebar has a size of 7~nm. c) Further zoomed image on the domain wall between the two domains at the location marked in c) with the same fitted lines. The shift of the SoD lattice between the two domains is visible by eye and most pronounced for the green and blue directions. The image has a total size of 7$\times$7~nm. d) The shifts of the top layer lattice between several pairs of domains with different electronic character overlaid on a schematic of a single SoD. The size of each cross indicates the uncertainty of the position. The blue and green circle represent the expected shift for AB and AC stacking respectively. The dotted black star corresponds to one possible orientation of the CDW (see main text). The choice of rotation is made so as to be in better agreement with the individual shifts. e) Schematic depiction of the proposed reason for the changes in stacking order. Left: the transition between insulating (red) and metallic (blue) phase across a top layer domain wall (DW) with the Ta atoms in the SoD drawn to illustrate the change in top layer stacking. The green arrow in the metallic (right) domain demonstrates the shift we calculate, as seen in d).
Right: view along the stacking direction with SoDs in the insulating and metallic phases of the top layer in solid colors and the SoDs of the second-to-top layer in dashed black. Note how the domain wall only occurs in the top layer while the periodicity of the second-to-top layer continues uninterrupted. The green arrow in the metallic domain again corresponds to the shift we calculate.}
\label{Fig:Shift_Analysis}
\end{figure*}

We will now focus on the immediate origins of the metallic behaviour in the mosaic phase, starting from the hypothesis that a change of stacking of the top layer to AB stacking due to the presence of domain walls in only one of the top two layers is responsible for this metalicity.
Using the assumption that the periodicity of the CDW in the second-to-top layer remains the same, that is to say, there is not coincidentally also a domain wall at the same location in the second-to-top layer, and knowing what stacking order is present in the normal insulating phase, one can reconstruct the stacking order in the metallic phase by extrapolating the lattice of one domain across the domain wall and comparing it to the lattice in a second domain.
When then comparing the shift between the two lattices with the offsets between Ta atoms in the regular atomic lattice, we can gain information about which shift has occurred across the domain wall and thus how the stacking changes.
Further technical details of the analysis process can be found in the supplementary material.

We apply this method to the image in Figure \ref{Fig:Shift_Analysis}a) which has several large domains with differing electronic character adjacent to each other, as seen from the visual contrast between the slightly darker metallic mosaic regions and the brighter insulating normal phase, which makes it suitable for our method of analysis.
We assume that the normal phase in this image is AA stacked, due to the contrast between the insulating and metallic domains which is the same as in figure \ref{Fig:SpatialOverview}b) which shows a large gap characteristic of this termination \cite{butler2020, zhang2022}.
Additionally, we have also observed a metallic mosaic on an AC stacked top layer, which shows a strikingly different contrast, as presented in the supplementary materials.
Figure \ref{Fig:Shift_Analysis}b) contains a zoom on two domains of different electronic character with fits applied along the three axes of the hexagonal CDW lattice represented by the coloured lines.
A further zoom onto the region around the domain wall is shown in Figure \ref{Fig:Shift_Analysis}c). Here, the change of periodicity between the two domains can be seen by eye when comparing the lines along the same lattice directions on either side of the domain wall.
Finally, the result of the analysis is presented in Figure \ref{Fig:Shift_Analysis}d).
The lattice shift between a total of nine sets of two adjacent domains each with different electronic character is indicated by the black crosses.
In essence, the black crosses represent the center positions of the SoDs of a metallic domain with respect to the center of the SoDs in an adjacent insulating domain.
The blue and green circles mark the distances of the inner Ta atoms of the SoDs (corresponding to an AB stacking order) and outer Ta atoms (corresponding to AC stacking) respectively while the red circled dot corresponds to AA stacking.
The obtained shifts thus agree with the shift leading to an AB stacking in the metallic domains, providing support for a non-bulk stacking related origin of the metallicity of the mosaic phase only from knowledge about the top layer.

Figure \ref{Fig:Shift_Analysis}e) presents a schematic explanation of how this shift occurs.
On the left, two layers of Ta atoms as well as the outline of the SoD CDW pattern can be seen.
On the left side of the domain wall, the top layer is stacked in the AA configuration and thus displays the expected insulating behaviour.
However, on the transition across the domain wall in the top layer, the phase of the top layer CDW changes with respect to that of the second-to-top layer since the domain wall is composed of Ta atoms not participating in the formation of any complete SoDs.
Again, we assume here that no domain wall is occurring in the second-to-top layer at the same location.
While domain walls in the second-to-top layer have been indirectly observed both in previous studies and in our own measurements \cite{butler2020,cho2016,ma2016}, there is no reason to assume that they consistently occur in the same pattern as the ones in the top layer.
As such, a spatial shift between the centres of the top and second-to-top layer CDW occurs, indicated by the green arrow, thus altering the stacking order.
The right panel provides a top-down view of the same situation with the second-to-top layer SoDs being indicated by dashed black lines.
In the insulating phase on the bottom, the SoDs overlap between the two layers whereas in the metallic phase, the AB stacking order occurs due to the broken periodicity in the top layer caused by the domain wall, while the periodicity of the bottom layer CDW is maintained as no domain wall occurs in that layer.
The shift which we have calculated for several domain walls in Figure \ref{Fig:Shift_Analysis}d) is again represented by a green arrow in the top domain.
Note that this same mechanism also allows for domain walls in the second-to-top layer, with no corresponding domain wall in the top layer, to change the stacking order within a single top layer domain.
An example of this occurring can be seen just above and to the right of the center of Figure \ref{Fig:SpatialOverview}b) where a visual contrast and a change in electronic behaviour occurs between the upper and lower parts of a single top layer domain.

We can further consider that the atomic lattice can only be aligned in two mirror symmetric ways relative to the CDW lattice.
These two potential orientations of the atomic lattice are normally indistinguishable from knowledge of only the CDW lattice.
However, as the structure of the CDW is not altered by lattice shifts, that is to say the shifted center of the SoD will nevertheless correspond to the location of a Ta atom, the obtained shifts should be located at the Ta locations of one of the two possible atomic lattices.
From our analysis, we find that the shifts indeed align reasonably well with one potential atomic lattice as indicated by the dotted black star in \ref{Fig:Shift_Analysis}e).
For reference, the second possible atomic lattice based on the CDW lattice would be rotated by thirty degrees around the center of the SoD, meaning the atoms would be located on the same circles in Figure \ref{Fig:Shift_Analysis}d) but centred between the atoms of the chosen lattice. This would correspond less well with our results.
Therefore, our analysis is potentially capable of extracting information about the underlying \textit{atomic} lattice from the behaviour of the CDW across domain walls, even without an image with atomic resolution.

We stress here that this natural mosaic phase has been observed across numerous samples, cleaves and macroscopically distinct locations on the surface of samples from different growth batches and produced by two different sample growers, each time accompanied by a closing of the electronic gap and a visual contrast between the two phases.
We have also observed the mosaic phase after various periods of time after cleaving the sample, including in the very first measurement.
Its appearance is therefore not likely to be related to sample ageing.
Furthermore, while the metallic mosaic phase is sometimes confined to small areas of 10$\times$10~nm$^2$, we also observe continuous mosaic regions extending over several hundreds of nanometres and covering a similar fraction of imaged surface area as the insulating normal phase.
When selecting different measurement locations on the surface of the same sample we have also noted that on multiple occasions, all measured locations within an area of $\sim$1~mm$^2$ on the sample surface contained a metallic mosaic while other areas of the sample only showed the normal insulating phase.
Additional STM and STS measurements demonstrating the above can be found in the supplementary material.
Given this, we attribute the appearance of this phase in our samples to a sample-intrinsic property which occurs with some significant frequency across TaS$_2$ samples.

We now turn to the question of why this mosaic phase appears in our samples without the need for external stimulation, in contrast to previous studies.
As mentioned above, the most immediately striking difference seen in our samples as compared to those in previous studies \cite{butler2020, ma2016, cho2016} is the larger amount of intrinsic defects in the CDW lattice, as seen for example in Figure \ref{Fig:Shift_Analysis}a).
To quantify this, we determined the number of defects in several STM images taken on samples which hosted the mosaic phase and found that an average of 2.9$\pm$0.4$\%$ of surface SoDs contained a defect in these images.
In comparison, this value is over twice that of the highest density of such defects found in the existing literature \cite{butler2020, ma2016, cho2016}.
Even more interestingly, recent investigations of Ti doped TaS$_2$ have also demonstrated the appearance of a metallic mosaic phase at low doping \cite{zhang2022}.
The density of Ti intercalation defects at the surface on which the mosaic phase is observed is 4.2$\%$ of surface SoDs, similar to our results, though from the bulk doping of 1$\%$ one would expect 12$\%$ of SoDs to show defects.
Note that the highest density of surface defects in our measurements remains below the minimum doping reported in ref. \cite{zhang2022} for which only changes to the CCDW transition temperature and minor changes to the electronic structure are observed.
This is further confirmed by the CCDW transition temperature in our samples which is the same as for pristine samples, see Figure \ref{Fig:Characterisation}d).
Thus, in spite of the density of defects in our samples being higher than in previous studies, this does not impact the macroscopic properties of our samples.
The defect density in our samples is also far removed from the doping level at which the insulating CDW phase is fully suppressed (8$\%$ Ti doping) \cite{zhang2022,gao2021}, in agreement with our observations.

One potential explanation for this behaviour is the pinning of domain walls at defects sites during the cooling from the near commensurate CDW phase to the CCDW phase.
Though possible, the domain walls which exist in the NCCDW phase are qualitatively different from those in the CCDW phase and it is not obvious how one would directly turn into the other while pinned \cite{park2019}.
There is a different possible non-pinning related explanation for the relationship between defects and the mosaic phase based on theoretical work aimed at explaining the origin of the pulse-induced mosaic in TaS$_2$ \cite{karpov2018}.
From simulations, it was found that charged defects introduced into a hexagonal lattice lead to the formation of networks of charged, one dimensional domain walls.
As such, it appears plausible that the increased density of defects at the surface of our samples causes the formation of a dense network of domain walls, giving rise to the characteristic appearance of the mosaic phase in the same manner as for intentional doping of the sample \cite{zhang2022}.
The resulting high density of domain walls increases the probability of a domain wall occurring only in one of the top two layers, thus leading to the appearance of the metallic phase in the mosaic due to a change in stacking of the top layer as shown above.
However, it is not clear why the majority of domains within the domain wall network would adopt this non-bulk stacking.
It may be possible that instead the formation of domain walls due to the effects of intrinsic defects upon cooling is favoured within regions of altered stacking.
These findings also provide insight into a potential mechanism of the creation of the pulse-induced mosaic phase, namely that the pulses create similar charged defects in the CDW lattice, which then cause the formation of domain walls as the lattice cools again \cite{stahl2020}.

In conclusion, we have shown the occurrence of a metallic mosaic phase at the surface of 1\textit{T}-TaS$_2$ without the need for external stimulation through laser or electrical pulses.
Our characterisation of this phase reveals that it behaves the same as the pulse induced phase, leading us to assume that they are the same and opening up the possibilities for study of this phase without the need for additional preparation steps.
In analogy with the behaviour of Ti doped TaS$_2$ samples and supported by previous model calculations, we suggest that the origin of the mosaic phase in our samples is found in a higher density of surface defects which lead to the formation of a network of domain walls.
Through further analysis of our measurements, we have demonstrated how interlayer stacking could plausibly be changed within the metallic surface domains due to the occurrence of domain walls in only one of the two top layers, which is made more likely by the presence of a large number of domain walls in the mosaic phase.
This mechanism can also serve as an explanation for the pulse-induced mosaic phase if one assumes that localised charged defects are created in the CDW lattice by the pulses.

\textit{Acknowledgements} 
B.S. and C.M. acknowledge support from the Swiss National Science Foundation Grant No. P00P2$\_$170597.

%\bibliography{UTaS2_Paper.bib}

%

\end{document}

\title{Supplementary Materials for\\Observation of metallic mosaic phase in 1\textit{T}-TaS$_2$ at equilibrium}

\author{B.~Salzmann}
\altaffiliation{Corresponding author.\\ bjoern.salzmann@unifr.ch}
\affiliation{D{\'e}partement de Physique and Fribourg Center for Nanomaterials, Universit{\'e} de Fribourg, CH-1700 Fribourg, Switzerland}

\author{E. Hujala}
\affiliation{D{\'e}partement de Physique and Fribourg Center for Nanomaterials, Universit{\'e} de Fribourg, CH-1700 Fribourg, Switzerland}
\affiliation{Lappeenranta-Lahti University of Technology LUT, FI-53850 Lappeenranta, Finland}

\author{C.~Witteveen}
\affiliation{Department of Chemistry, University of Zurich, CH-8057 Zürich, Switzerland}
\affiliation{Department of Quantum Matter Physics, University of Geneva, CH-1211 Geneva, Switzerland}

\author{B.~Hildebrand}
\affiliation{D{\'e}partement de Physique and Fribourg Center for Nanomaterials, Universit{\'e} de Fribourg, CH-1700 Fribourg, Switzerland}

\author{H.~Berger}
\affiliation{Institut de Physique des Nanostructures, {\'E}cole Polytechnique F{\'e}d{\'e}rale de Lausanne (EPFL), CH-1015 Lausanne, Switzerland}

\author{F.O.~von Rohr}
\affiliation{Department of Chemistry, University of Zurich, CH-8057 Zürich, Switzerland}
\affiliation{Department of Quantum Matter Physics, University of Geneva, CH-1211 Geneva, Switzerland}

\author{C.W.~Nicholson}
\affiliation{D{\'e}partement de Physique and Fribourg Center for Nanomaterials, Universit{\'e} de Fribourg, CH-1700 Fribourg, Switzerland}
\affiliation{Fritz Haber Institute of the Max Planck Society, D-14195, Berlin, Germany}

\author{C.~Monney}
\affiliation{D{\'e}partement de Physique and Fribourg Center for Nanomaterials, Universit{\'e} de Fribourg, CH-1700 Fribourg, Switzerland}

\date{\today}
\maketitle

\subsection*{Additional Characterisation Measurements}

\begin{figure*}
\includegraphics[width=0.8\linewidth]{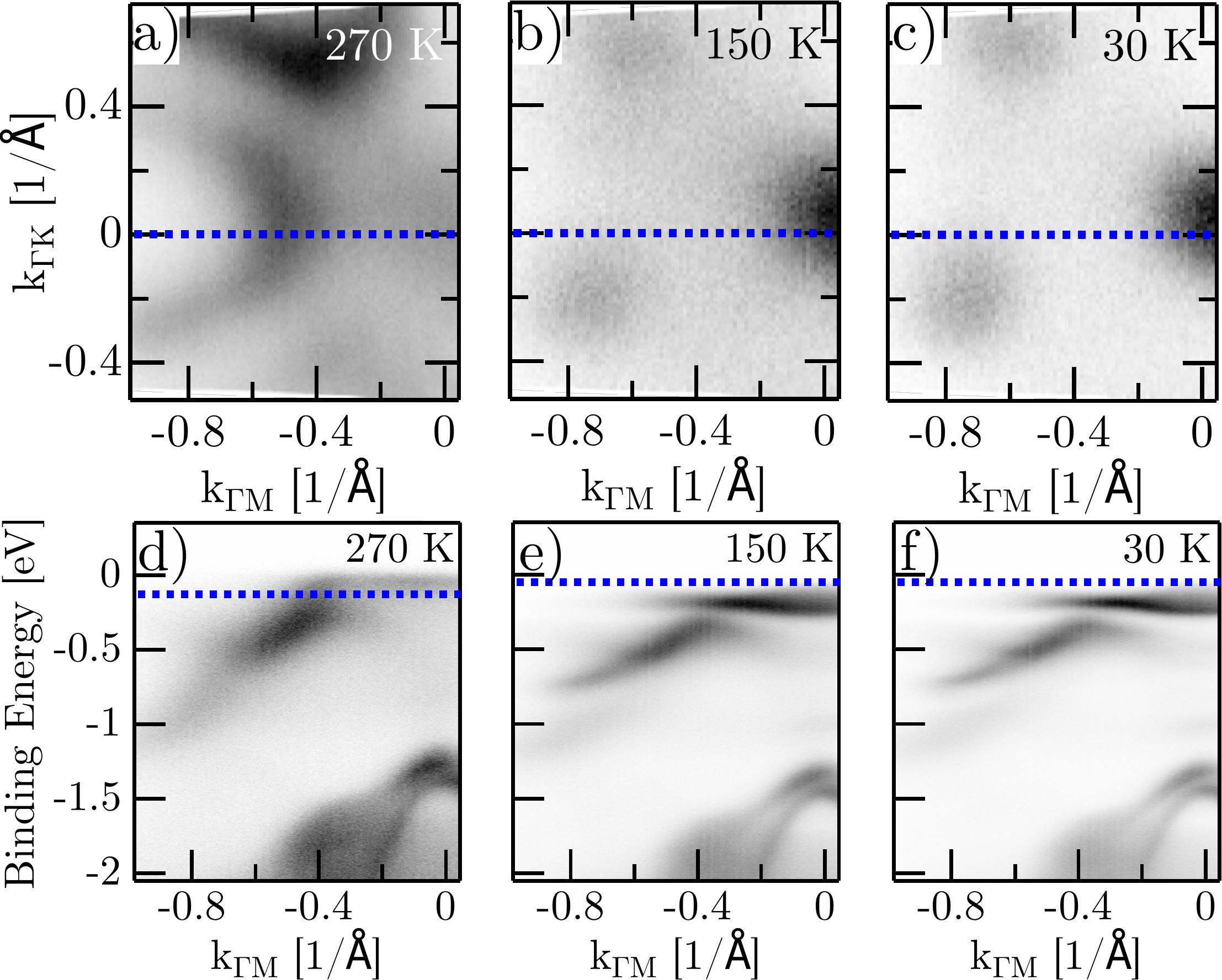} 
\caption{Additional ARPES measurements in the near commensurate CDW and CCDW phases. a)-c) Constant energy surfaces at three different temperatures. d)-e) Momentum cuts along the $\bar{\Gamma}\bar{\mathrm{M}}$ direction at the locations indicated by the blue dashed lines in a) to c). Blue dashed lines indicate the energy of the surfaces in a)-c).}
\label{Fig:Additional_ARPES}
\end{figure*}

As mentioned in the main text, we have performed more in-depth characterisation measurements on our samples in order to verify that they do indeed behave like 1\textit{T}-TaS$_2$.
Figure \ref{Fig:Additional_ARPES} presents additional ARPES measurements at different temperatures taken with 21.2~eV photon energy as in the main text.
Figures \ref{Fig:Additional_ARPES}a)-c) contain constant energy surfaces obtained at temperatures above and below the CCDW transition in good agreement with previous measurements \cite{perfetti2005,ngankeu2017,wang2020}.
Note that the (0,0) position corresponding to the $\bar{\Gamma}$ point is to the right of the image in order to show the characteristic features of the Fermi surface.
Figures \ref{Fig:Additional_ARPES}d)-f) show spectra taken at the k position indicated in a)-c) again showing the expected changes across the CCDW transition.
In all cases, we observe only minor changes between the measurements taken at 150~K and 30~K related to less thermal broadening of the features at lower temperatures.

\begin{figure*}
\includegraphics[width=0.8\linewidth]{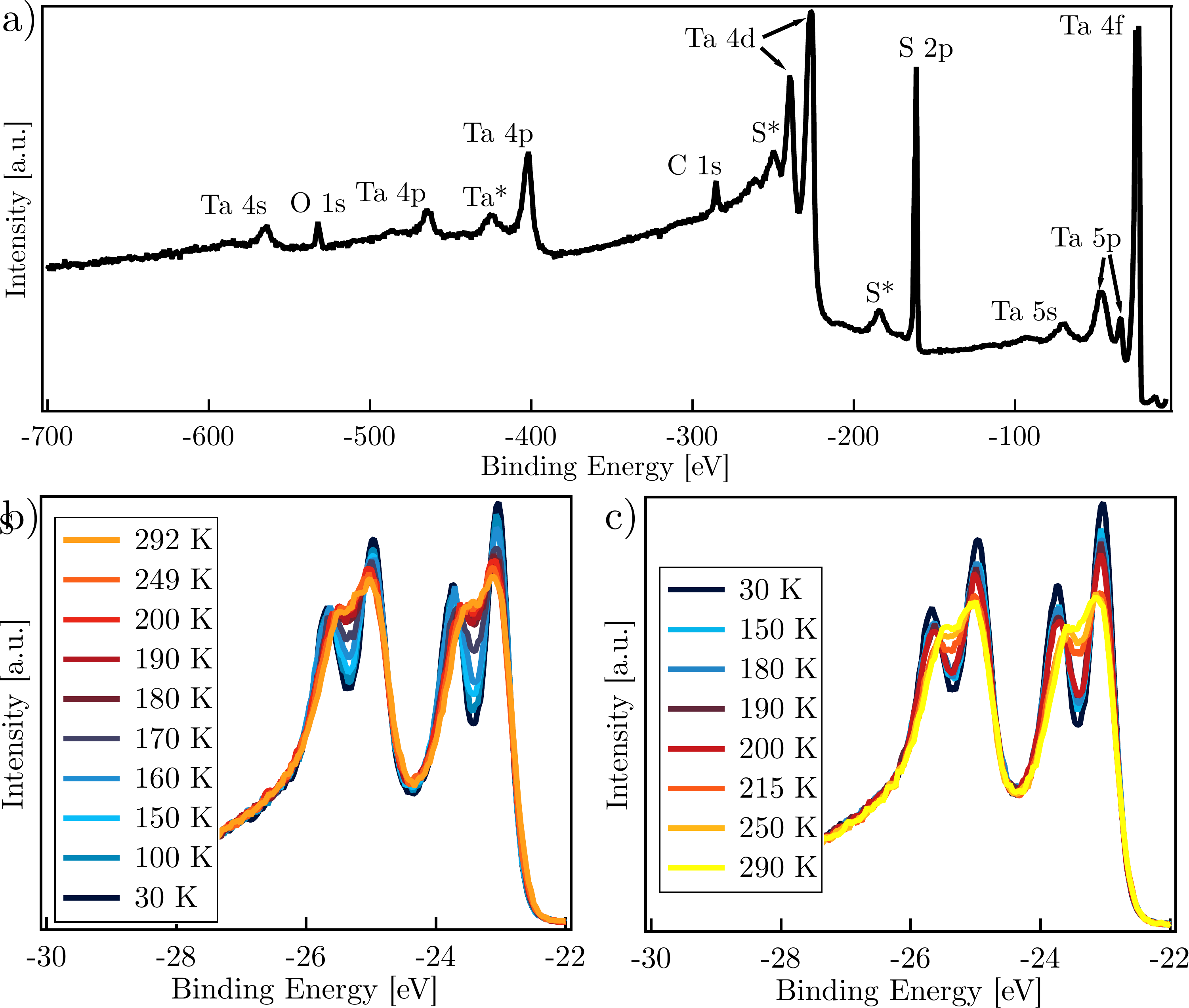} 
\caption{a) XPS spectrum with the principal peaks marked. Stars indicate peaks attributed to plasmon losses. b) XPS spectra of the Ta 4f core levels upon cooling from room temperature showing the transition to the CCDW phase between 170~K and 160~K. c) XPS spectra of the same core levels upon warming. A phase transition occurs between 200~K and 215~K.}
\label{Fig:Additional_XPS}
\end{figure*}

Figure \ref{Fig:Additional_XPS}a) displays an XPS spectrum over a larger energy range where we have identified the core levels of the Ta and S atoms.
While small amounts of common contaminants such as oxygen and carbon are present on the sample, there does not appear to be any contamination by foreign elements that would call into question to chemical composition of our sample.
Figure \ref{Fig:Additional_XPS}b) contains the evolution of the Ta 4f peaks upon cooling with the transition to the CCDW phase, indicated by a sudden increase of the splitting of both double peaks, occurring between 170~K and 160~K, in agreement with resistivity and ARPES data.
After this transition, again no change in behaviour is observed down to 30~K.
Figure \ref{Fig:Additional_XPS}c) shows the warming up part of the temperature series beginning in b), obtained directly after the measurements in b) on the same cleaved surface.

\subsection*{Metallic Character of the Mosaic Phase}

\begin{figure*}[tb]
\includegraphics[width=0.8\linewidth]{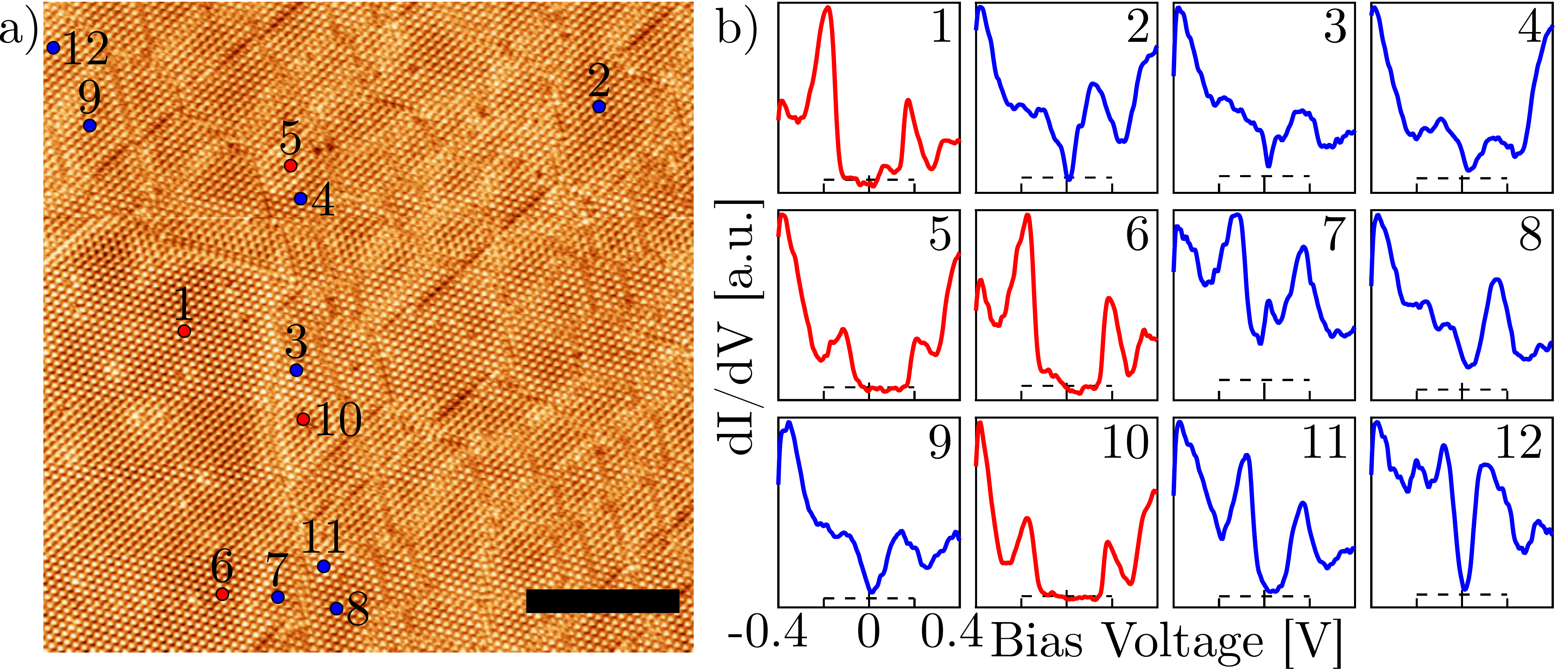} 
\caption{Larger image of the region shown in figure 2b) in the main text. The scalebar has a size of 20~nm. b) STS spectra taken at various points marked by the numbered circles in a). Spectra shown in red are insulating while spectra shown in blue are metallic. As stated in the main text, the visual contrast between different domains perfectly predicts the electronic character of the domains even when the domain wall is burried, as is e.g. the case between points 4 and 5.}
\label{Fig:Mosaic_Characterisation}
\end{figure*}

In the following, we present additional STS measurements taken in the normal and mosaic phase in the region shown in figure 2b) of the main text to demonstrate how the mosaic phase and the visual contrast seen in STM images relate to the electronic character of different domains. 
Figure \ref{Fig:Mosaic_Characterisation}a) shows a larger topographic image of the area with the region shown in the main text located in the bottom left.
The numbered dots correspond to individual STS measurements, with red and blue dots corresponding to insulating and metallic spectra respectively.
Locations for STS measurements were selected by hand to lie at the center of SoDs as well as possible and not within any domain wall so as to gain consistent information about the character of the actual domains.
Note that the two measurements from the main text are not included in this figure.
Figure \ref{Fig:Mosaic_Characterisation}b) shows the spectra of all twelve points marked in \ref{Fig:Mosaic_Characterisation}a).
The bias voltage range is the same for all spectra.
Each spectrum corresponds to the average of ten individual voltage sweeps performed in succession at the same location.
The dashed lines show the position of 0 dI/dV.
While the spectra within the two different regions are generally similar, there is still quite some variation.
This can be attributed to several different causes.
Firstly, residual electronic and vibrational noise is only partially mitigated by our choice of number of measurements.
Secondly, small changes of the tip electronic state are likely to occur during voltage sweeps or tip movement between different measurement locations, resulting in differences in the measured currents.
Finally, as we can only manually select the locations of the spectra based on the topographic image, the exact position within the SoD we probe is not necessarily consistent, which is likely repsonsible for e.g. the variation in the height of the peaks on either side of the gap in the insulating phase.
We note here that while the vast majority of the mosaic phase is metallic, the most reliable identifier of the electronic structure lies in the visual contrast between insulating and metallic regions in the topographic image.
For example, spectrum five is clearly taken within the mosaic phase, yet is obviously insulating.
However, when looking at the location of spectrum five, one can see that it is in a small region brighter than its surroundings which are metallic as seen in spectrum four, similar to how the large non-mosaic insulating region in the bottom left is brighter than the surrounding metallic mosaic.
Indeed, as mentioned in the main text, while there is no visible domain wall between locations five and four, there is a clear visual contrast between their respective parts of the domain, which can be attributed to a domain wall in the second to top layer which causes a change in stacking.
The same is true for the difference between spectra 3 and 10 and 6 and 7, where spectrum 7 was taken in a small, darker part at the edge of the otherwise insulating large domain and indeed shows clearly metallic behaviour.
This visual contrast is likely due to the increased density of states due to the peak around -0.2V in the insulating phase.

\subsection*{Spatial Extent of the Mosaic Phase}

\begin{figure*}[tb]
\includegraphics[width=0.6\linewidth]{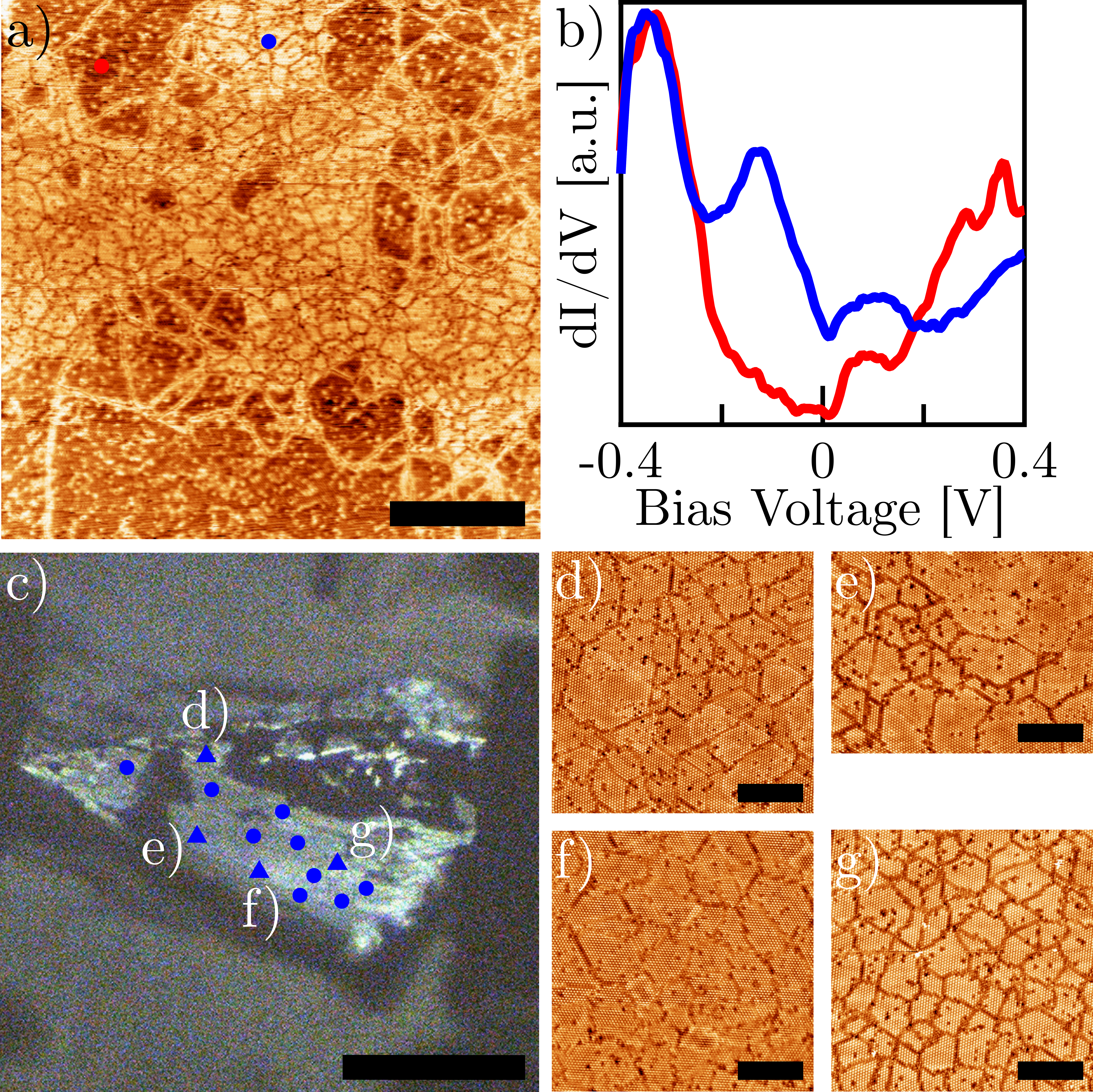} 
\caption{a) Large scale STM image showing the spatial extent of the mosaic phase. The scalebar has a size of 100~nm. Note the strong contrast between the mosaic and normal regions caused by their different electronic character. b) STS spectra taken at the correspondingly colored points in a). The mosaic region again appears metallic. Note that the normal phase here is still insulating, but is likely AC stacked. c) Image of a typical TaS$_2$ sample in the STM showing the cleave. The scalebar has a size of 2mm. The circles and triangles corresponds to the macroscopic location of the STM measurements taken on this cleave. Each location showed the mosaic phase. d)-g) STM images obtained at the correspondingly labelled triangles in c), all showing the mosaic phase either across the entire or parts of the image. Each scalebar has a size of 25~nm.}
\label{Fig:Spatial_Extent}
\end{figure*}

As mentioned in the main text, we have found that the mosaic phase has on several occasions been found to extend over distances larger than 400~nm.
One such image with a size of 400x400~nm$^2$ can be seen in figure \ref{Fig:Spatial_Extent}a).
Note that here, the mosaic phase appears much brighter than the normal phase which can be explained by the STS spectra taken at the marked locations with the same parameters as described above.
These spectra are presented in figure \ref{Fig:Spatial_Extent}b) and again confirm the metallic character of the mosaic phase.
However, the spectrum in the normal phase appears more similar to an AC stacked surface, due to its smaller gap (see refs. \cite{butler2020,wu2021} for comparison).
This explains why we see an inversion of which phase appears brighter in the topographic image.

By studying several locations on the same cleaved surface, we can gain further information about the spatial distribution of the mosaic phase.
Figure \ref{Fig:Spatial_Extent}c) shows a TaS$_2$ sample as seen during measurements in our STM.
Note that while the image is about 5~mm across horizontally, it is strongly foreshortened vertically due to the flat angle of the camera.
The sample can be identified by the bright silver reflection as opposed to the duller grey sample holder below. 
Blue markers (both circles and triangles) show the approximate macroscopic locations of where measurements were performed on this cleave.
At each of these points, we have taken at least three topographic images at different microscopic locations.
For each of these sets of measurements, at least one and frequently every single one showed the mosaic phase, confirming that the mosaic phase is not a localised phenomenon but may be observed consistently across distances of several millimetres on a sample.
Figures \ref{Fig:Spatial_Extent}d)-g) present four such topographic images taken at the locations marked by triangles and labelled in c).
The selected locations were chosen to be spread widely across the surface of the sample.

\subsection*{Algorithm for STM image analysis}

We now present a more detailed explanation of how we perform the analysis of the lattice shift across domain walls.
Image analysis of the original STM data is based on pattern recognition methods, which are currently applied in multiple engineering applications \cite{Hujala2018, Patel2022}. The basic principles of the algorithm, written in MATLAB \cite{Matlab2020}, are presented as a gif-animation which consists of six steps A to F. The algorithm reads the original STM data (Step A). Edge detection detects placement and thickness of the domain walls while different domains are labeled as shown in Step B of the animation. Centres of the SoDs are evaluated using Otsu's method \cite{Otsu1979}. Combining this with the labels of the domains, center points of each SoD of all domains are known as shown in different coloured asterisks in Step C.  At this point, the algorithm also recognizes the defects inside the domains. The least square method is used to fit vectors to centroid data for each of the three directions shown in yellow, cyan and magenta lines in Step D. In order to compare adjacent vectors between two domains properly, an average slope for each direction is calculated ($\bar{a}_1$, $\bar{a}_2$, $\bar{a}_3$) and new vectors are fitted (Step E). The final step E connects consecutive vectors over the defects. We also extract a list of intercept values for each individual fitted line in all three directions, separately for both domains. More detailed description of the algorithm will be presented in a future publication. 

Starting from the lists of the intercepts of individual rows of SoDs in both domains, we proceed with the calculation of the shift between domains.
The general idea behind the calculation is relatively simple: from the knowledge of the locations of each row of SoDs in either domain (as given by the extracted intercepts), we first compute the average difference in intercepts between the domains along each of the three directions ($\bar{s}_1$, $\bar{s}_2$, $\bar{s}_3$) corresponding to the three equivalent directions of the hexagonal CDW lattice, essentially obtaining three new lines given by the equations

\begin{equation}
y = \bar{a}_i \cdot x + \bar{s}_i, \qquad i \in \{1, 2, 3\}
\end{equation}

From the crossing point of these lines we find a location which represents the average location of the SoD centres of the second domain with respect to those of the first domain which now correspond to the point (0,0) making it trivial to extract the exact size and direction of the shift.
This corresponds to the black crosses shown in figure 3d) of the main text where again, the (0,0) coordinate corresponds to the centre of the SoDs in the reference domain.
Note that the shown calculated SoDs have been shifted into the first unit cell of the CDW lattice for ease of visualisation but the absolute values of shifts can exceed the size of a single SoD.

Complications arise from the fact that the STM image is subtly distorted due to an unequal piezo response as well as small amounts of drift over the course of the measurement.
These distortions manifest themselves in two ways.
Firstly, the slopes of the rows of SoDs are not consistent across the image.
We address this issue by calculating the average slopes locally only for SoDs in the two domains of interest for each point.
Secondly, the distances between rows within a single domain depend on the direction of the row and this dependence is significant compared to the variations between individual inter-row distances along the same direction.
Quantitatively, our STM image shows an inter-row distance which is between 10\% and 5\% lower than literature values, depending on the direction which we attribute to the effects of ageing of the piezo actuators.
We address this issue by normalising each distance $\bar{s}_i$ with the ratio between the average distance along this direction and the expected lattice spacing.
To further lower potentially accumulating errors, we consider only nearest neighbour pairs of rows when calculating the differences in intercepts.
Because of this, depending on the relative orientation of the two domains, it is possible that not all three directions contain any nearest neighbour pair of rows in which case we use only the crossing point of the remaining two directions.

\begin{figure*}
\includegraphics[width=0.4\linewidth]{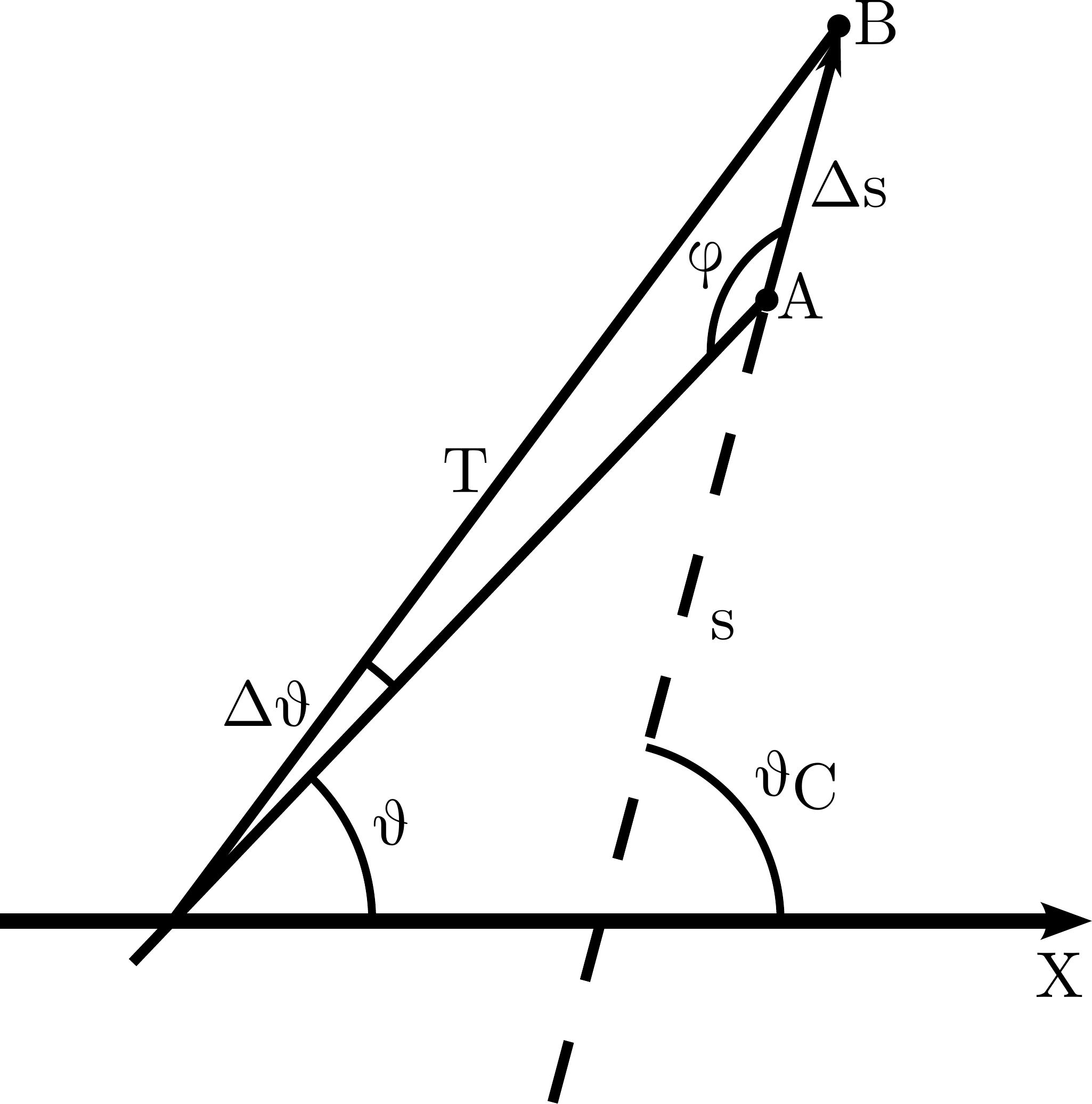} 
\caption{Sketch showing the geometry of our slope adjustment. When the SoD location is shifted along the direction $\vartheta_C$ by an amount $\Delta s$ (moving from point A to point B), the angle of the slope $\vartheta$ has to be adjusted by an amount $\Delta \vartheta$.}
\label{Fig:SlopeAdjust_Sketch}
\end{figure*}

However, correcting only for distance while keeping the slopes the same can lead to a misalignment of the crossing points which we correct for by adjusting the slopes depending on the amount of correction for the distance.
Figure \ref{Fig:SlopeAdjust_Sketch} shows the geometry of the problem.
If we shift a SoD location from point A to point B along a direction under the x axis of $\vartheta_C$, i.e. when we correct for the lattice distortion of the inter-row distances of rows perpendicular to $\vartheta_C$, in order to keep the three crossing points of the three different row directions the same, we have to adjust the slope in the direction $\vartheta$ by an angle of $\Delta \vartheta$ such that this row now passes through B instead of A.
Geometrically, we can calculate the angle $\varphi$ as

\begin{equation}
\varphi = \frac{360^\circ - 2\cdot (\vartheta- \vartheta_C)}{2}
\end{equation}

Then, applying the law of sines, we obtain the adjustment angle $\Delta \vartheta$ as

\begin{equation}
\Delta \vartheta = \mathrm{arcsin}\left(\frac{\mathrm{sin}(\varphi)}{T}\Delta s\right)
\end{equation}

where we have used that the distance between the SoDs after the correction, $T$, is the true lattice constant while $\Delta s$ can be calculated from the correction factor we apply to the distance $s$.
Note that this correction is applied twice to each slope, once for each non-perpendicular inter-row distance.
We find however that the influence on our final results from this correction step is only minor.

%\bibliographystyle{naturemag}
%\bibliography{UTaS2_Paper}